\documentstyle[preprint,eqsecnum,aps,prd]{revtex}
\begin{document}
\draft
\preprint{\parbox{2.2in}{FERMILAB--PUB--94/032--T\\hep-ph/9402210\\[12pt]}}
\title{Mesons with Beauty and Charm: Spectroscopy}
\author{Estia J. Eichten\thanks{Internet address: eichten@fnal.gov}
and Chris Quigg\thanks{Internet address: quigg@fnal.gov}}
\address{Theoretical Physics Department\\
Fermi National Accelerator Laboratory\\
P.O. Box 500, Batavia, Illinois 60510}
\date{\today}
\maketitle
\begin{abstract}
Applying knowledge of the interaction between heavy quarks derived from
the study of $c\overline{c}$ and $b\overline{b}$ bound states, we
calculate the spectrum of $c\overline{b}$ mesons.  We compute transition
rates for the electromagnetic and hadronic cascades that lead from excited
states to the $^1\text{S}_0$ ground state, and briefly consider the
prospects for experimental observation of the spectrum.
\end{abstract}
\section{Introduction}
The copious production of $b$ quarks in $Z^0$ decays at the Large
Electron-Positron collider (LEP) and in 1.8-TeV proton-antiproton
collisions at the Fermilab Tevatron opens for study the rich spectroscopy
of mesons and baryons beyond $B_u^+$ and $B_d^0$.  In addition to $B_s^0$
and $\Lambda_b^0$, which have already been widely discussed, a particularly
interesting case is the spectrum of $c\overline{b}$ states and its ground
state, the $B_c^+$ meson \cite{PDG}.

Even more than their counterparts in the $J\!/\!\psi$ and $\Upsilon$ families,
the $c\overline{b}$ states that lie below the ($BD$) threshold for decay into a
pair of heavy-flavored mesons are stable against strong decay, for
they cannot annihilate into gluons.  Their allowed decays, by E1 or M1
transitions or by hadronic cascades, lead to total widths that are less
than a few hundred keV.  All decay chains ultimately reach the $^1$S$_0$
ground state $B_c$, which decays weakly.  It may be possible, in time, to
map out the excitation spectrum by observing photons
or light hadrons in coincidence with a
prominent decay of the $B_c$ \cite{CDFchi}. This would
test our understanding of the force between heavy quarks.

The weak decays of the $c\overline{b}$ ground state will be of particular
interest because the influence of the strong interaction can be estimated
reliably \cite{eqweak}.  The deep binding of the heavy quarks within the
$B_c$ means that
the spectator picture is misleading.  Taking proper account of binding
energy, we expect a rather long lifetime
that implies easily observable secondary vertices.  The deep binding also
affects the $B_c$ branching fractions and leads us to expect that final
states involving $\psi$ will be prominent.  The modes $\psi \pi^+$, $\psi
a_1^+$, $\psi\rho^+$, $\psi D_s^+$, and $\psi \ell^+ \nu_\ell$ will serve
to identify $B_c$ mesons and determine the $B_c$ mass and lifetime.

In this Article, we present a comprehensive portrait of the spectroscopy of
the $B_c$ meson and its long-lived excited states.  In Section II, we
estimate the mass of the $B_c$ in the framework of nonrelativistic
quarkonium quantum mechanics and calculate the spectrum of $c\overline{b}$
states in detail.  In Section III, we compute rates for the prominent
radiative decays
of the excited states and estimate rates and spectra of the hadronic
cascades $(c\overline{b})_i\rightarrow \pi\pi+(c\overline{b})_f$ and
$(c\overline{b})_i\rightarrow \eta+(c\overline{b})_f$.  Using this
information, we outline a strategy for partially reconstructing the
$c\overline{b}$ spectrum.  A brief summary appears in
Section IV.

\section{The Spectrum of $B_{\lowercase{c}}$ States}
\label{sec: spectrum}
\subsection{The Mass of $B_{\lowercase{c}}$}
Both in mass and in size, the mesons with beauty and charm are intermediate
between the $c\overline{c}$ and $b\overline{b}$ states.  Estimates of the
$B_c$ mass can, consequently, be tied to what is known about the charmonium
and $\Upsilon$ families.  To predict the full spectrum and properties of
$c\overline{b}$ states, we rely on the nonrelativistic potential-model
description of quarkonium levels.  The interquark potential is known rather
accurately in the region of space important for the $J\!/\!\psi$ and
$\Upsilon$ families \cite{Cornell,QR,KRQ}, which spans the distances
important for $c\overline{b}$ levels.  This region lies between the
short-distance Coulombic and long-distance linear behavior expected in QCD.
 We consider four functional forms for the potential that give reasonable
accounts of the $c\overline{c}$ and $b\overline{b}$ spectra: the QCD-motivated
potential \cite{Rich} given by Buchm\"{u}ller and
Tye \cite{BT}, with
\begin{eqnarray}
m_c = 1.48\text{ GeV}\!/\!c^2 & \;\;\; &m_b = 4.88\text{ GeV}\!/\!c^2\;\;;
\end{eqnarray}
a power-law potential \cite{Martin},
\begin{equation}
	V(r) = -8.064\text{ GeV} + (6.898\text{ GeV})(r\cdot\text{1 GeV})^{0.1}
	 \;\;, \label{Martin}
\end{equation}
with
\begin{eqnarray}
m_c = 1.8\text{ GeV}\!/\!c^2 & \;\;\; & m_b = 5.174\text{ GeV}\!/\!c^2\;\;;
\end{eqnarray}
a logarithmic potential \cite{Log},
\begin{equation}
	V(r) = -0.6635\text{ GeV} + (0.733\text{ GeV}) \log{(r\cdot\text{1
	GeV})}\;\;,
\end{equation}
with
\begin{eqnarray}
m_c = 1.5\text{ GeV}\!/\!c^2 & \;\;\; & m_b = 4.906\text{ GeV}\!/\!c^2\;\;;
\end{eqnarray}
and a Coulomb-plus-linear potential (the ``Cornell potential'')
\cite{Cornell},
\begin{equation}
	V(r)=-\frac{\kappa}{r} + \frac{r}{a^2}\;\;,
	\label{cornellpot}
\end{equation}
with
\begin{eqnarray}
m_c = 1.84\text{ GeV}\!/\!c^2 & \;\;\; & m_b = 5.18\text{ GeV}\!/\!c^2 \\
\kappa=0.52 & \;\;\; & a=2.34\text{ GeV}^{-1}\;\;.
\end{eqnarray}

We solve the Schr\"{o}dinger equation for each of the potentials to
determine the position of the 1S center of gravity for $c\overline{c}$,
$c\overline{b}$, and $b\overline{b}$.  The $^3$S$_1$ -- $^1$S$_0$
splitting of the $i\bar{\jmath}$ ground state is given by
\begin{equation}
M(^3\text{S}_1) - M(^1\text{S}_0) = \frac{32\pi\alpha_s |\Psi(0)|^2}{9m_i m_j}
\;\;\; .
\end{equation}
The hyperfine splitting observed in the charmonium family \cite{PDG},
\begin{equation}
M(J\!/\!\psi) - M(\eta_c) = 117\text{ MeV}\!/\!c^2 \;\;\;, \label{psisplit}
\end{equation}
fixes the strong coupling constant for each potential.
We neglect the variation of $\alpha_s$ with momentum and scale the splitting of
$c\overline{b}$ and $b\overline{b}$ from the charmonium value
(\ref{psisplit}).
The resulting values of vector and pseudoscalar masses are presented in
Table \ref{Mass1S}.  Predictions for the $c\overline{b}$ ground-state
masses depend little on the potential.  The $B_c$ and $B_c^*$ masses and
splitting lie within the ranges quoted by Kwong and Rosner \cite{KR} in
their survey of techniques for estimating the masses of the $c\overline{b}$
ground state.  They find
\begin{equation}
6.194\text{ GeV}\!/\!c^2 \alt M_{B_c} \alt 6.292\text{ GeV}\!/\!c^2 \;\;\; ,
\end{equation}
and
\begin{equation}
6.284\text{ GeV}\!/\!c^2 \alt M_{B_c^*} \alt 6.357\text{ GeV}\!/\!c^2 \;\;\; ,
\end{equation}
with
\begin{equation}
65\text{ MeV}\!/\!c^2 \alt M_{B_c^*} - M_{B_c} \alt 90\text{ MeV}\!/\!c^2\;\;\;
{}.
\end{equation}
We take
\begin{equation}
M_{B_c} = 6.258 \pm 0.020\text{ GeV}\!/\!c^2
\end{equation}
as our best guess for the interval in which $B_c$ will be found
\cite{BBZ}.

We shall adopt the Buchm\"{u}ller-Tye
potential \cite{BT}
for the detailed calculations that follow, because it has the correct
two-loop short-distance behavior in perturbative QCD.

\subsection{Excited States}
The interaction energies of a heavy quark-antiquark system probe the basic
dynamics of the strong interaction.  The gross structure of the
quarkonium spectrum reflects the shape of the interquark potential.
In the absence of light quarks, the
static energy explicitly exhibits linear confinement at large distance.
Further insight can be obtained by studying the spin-dependent forces,
which distinguish
the electric and magnetic parts of the interactions.  Within the
framework of quantum chromodynamics, the nature of the spin-dependent
forces was first studied nonperturbatively by Eichten and Feinberg
\cite{EF1,EF2}.  Gromes \cite{Gromes} subsequently added an important
constraint that arises from boost-invariance of the QCD forms \cite{Peskin}.
One-loop perturbative QCD calculations for the spin-dependent interactions
in a meson composed of two different heavy quarks have also been carried out
\cite{Pqcd,TypP,NPT}.

The spin-dependent contributions to the $c\overline{b}$
masses may be written as
\begin{equation}
\Delta = \sum_{k=1}^4 T_k \;\;\; ,
\end{equation}
where the individual terms are
\begin{eqnarray}
T_1 & = &
\frac{\langle\vec{L}\cdot\vec{s}_i\rangle}{2m_i^2}\tilde{T}_1(m_i,m_j)
+ \frac{\langle\vec{L}\cdot\vec{s}_j\rangle}{2m_j^2}
\tilde{T}_1(m_j,m_i) \nonumber \\
T_2 & = & \frac{\langle\vec{L}\cdot\vec{s}_i\rangle}{m_i m_j}
\tilde{T}_2(m_i,m_j) + \frac{\langle\vec{L}\cdot\vec{s}_j\rangle}{m_i m_j}
\tilde{T}_2(m_j,m_i)  \label{T1to4}\\
T_3 & = & \frac{\langle\vec{s}_i\cdot\vec{s}_j\rangle}{m_im_j}
\tilde{T}_3(m_i,m_j) \nonumber \\
T_4 & = & \frac{\langle S_{ij}\rangle}{m_im_j} \tilde{T}_4(m_i,m_j) \;\;,
\nonumber
\end{eqnarray}
and the tensor operator is
\begin{equation}
S_{ij} = 4\left[3(\vec{s}_i\cdot\hat{n})(\vec{s}_j\cdot\hat{n})-
\vec{s}_i\cdot\vec{s}_j\right]\;\;. \label{sij}
\end{equation}
In Eq.  (\ref{T1to4}) and (\ref{sij}), $\vec{s}_i$ and $\vec{s}_j$ are the
spins of the heavy quarks, $\vec{L}$ is the orbital angular momentum of
quark and antiquark in the bound state, and $\hat{n}$ is an arbitrary unit
vector.  The total spin is $\vec{S}=\vec{s}_i+\vec{s}_j$.

The leading contributions to the $\tilde{T}_k$ have no explicit dependence
on the quark masses.  Assuming that the magnetic interactions are
short-range ($\propto\langle r^{-3} \rangle$) and thus can be calculated in
perturbation theory, we have
\begin{eqnarray}
\tilde{T}_1(m_i,m_j) & = & -\left\langle\frac{1}{r}\frac{dV}{dR}\right\rangle
   + 2\tilde{T}_2(m_i,m_j) \nonumber \\
\tilde{T}_2(m_i,m_j) & = & \frac{4\alpha_s}{3}\langle r^{-3} \rangle \\
\tilde{T}_3(m_i,m_j) & = & \frac{32\pi\alpha_s}{9}|\Psi (0)|^2 \nonumber \\
\tilde{T}_4(m_i,m_j) & = & \frac{\alpha_s}{3}\langle r^{-3} \rangle
\;.\nonumber
\end{eqnarray}
The connection between $\tilde{T}_1$ and $\tilde{T}_2$ is Gromes's
general relation; the other equations reflect the stated approximations.

For quarkonium systems composed of equal-mass heavy quarks, the total
spin $S$ is a good
quantum number and $LS$ coupling leads to the familiar classification of
states as $^{2S+1}L_J$, where $\vec{J}=\vec{L}+\vec{S}$ \cite{Dwave}.
The calculated spectra are compared with experiment in Table
\ref{charmonium} (for the $\psi$ family) and Table \ref{upsilons}
(for the $\Upsilon$ family).  Overall, the agreement is satisfactory.
Typical deviations in the charmonium system are less than about $30\text{
MeV}$; deviations in the upsilon system are somewhat smaller.  The
differences between calculated and observed spectra suggest that the
excitation energies in the $c\bar{b}$ system can be predicted within a
few tens of MeV.

The leptonic decay rate of a neutral ($Q\bar{Q}$)
vector meson $V^0$ is
related to the Schr\"{o}dinger wave function through \cite{vrw,QCDRC}
\begin{equation}
	\Gamma(V^0\rightarrow e^+ e^-) = \frac{16\pi N_c \alpha^2 e_Q^2}{3}
	\frac{|\Psi(0)|^2}{M_V^2} \left(1-{\displaystyle
	\frac{16\alpha_s}{3\pi}}\right)\;\;,
	\label{vrwf}
\end{equation}
where $N_c=3$ is the number of quark colors, $e_Q$ is the heavy-quark
charge, and $M_V$ is the mass of the vector meson.  The resulting leptonic
widths, evaluated without QCD corrections, are tabulated in Tables
\ref{charmonium} and \ref{upsilons}.  Within each family, the leptonic
widths are predicted in proper proportions, but are larger than the
observed values.  The QCD correction reduces the magnitudes
significantly; the amount of this reduction is somewhat uncertain,
because the first term in the perturbation expansion is large
\cite{correct}.

For unequal-mass quarks, it is more convenient to construct the mass
eigenstates by $jj$ coupling, first coupling $\vec{L}+\vec{s}_c=\vec{J}_c$
and then adding the spin of the heavier quark,
$\vec{s}_b+\vec{J}_c=\vec{J}$.  The level shifts $\Delta^{(J)}$ for the
$L=1$ states
with $(J_c=\case 3/2, J=2)$ and $(J_c=\case 1/2, J=0)$ are
\begin{eqnarray}
	\Delta^{(2)} & = & \left(\frac{1}{4m_c^2}+\frac{1}{4m_b^2}\right)
	\tilde{T}_1 + \frac{1}{m_bm_c}\tilde{T}_2 -\frac{2}{5m_bm_c}\tilde{T}_4
	\label{shift1} \\
	\Delta^{(0)} & = & -\left(\frac{1}{2m_c^2}+\frac{1}{2m_b^2}\right)
	\tilde{T}_1 -\frac{2}{m_bm_c}\tilde{T}_2 -\frac{4}{m_bm_c}\tilde{T}_4
	\nonumber\;\;.
\end{eqnarray}
For a given principal quantum number, the
two $(L=1,J=1)$ $c\overline{b}$ states with $J_c=\frac{1}{2}$ and
$\frac{3}{2}$ are mixed in general.  The elements of the mixing matrix are
\begin{displaymath}
	\Delta_{\frac{3}{2}\frac{3}{2}}^{(1)} =
\left(\frac{1}{4m_c^2}-\frac{5}{12m_b^2}\right) \tilde{T}_1 -
\frac{1}{3m_bm_c}\tilde{T}_2 + \frac{2}{3m_bm_c}\tilde{T}_4
\end{displaymath}
\begin{equation}
\Delta_{\frac{3}{2}\frac{1}{2}}^{(1)} =
\Delta_{\frac{1}{2}\frac{3}{2}}^{(1)}  =
-\frac{\sqrt{2}}{6m_b^2}\tilde{T}_1 - \frac{\sqrt{2}}{3m_bm_c}\tilde{T}_2
+\frac{2\sqrt{2}}{3m_bm_c}\tilde{T}_4  \label{unequal}
\end{equation}
\begin{displaymath}
\Delta_{\frac{1}{2}\frac{1}{2}}^{(1)} =
\left(-\frac{1}{2m_c^2}+\frac{1}{6m_b^2}\right) \tilde{T}_1 -
\frac{2}{3m_bm_c}\tilde{T}_2 + \frac{4}{3m_bm_c}\tilde{T}_4 \nonumber\;\;.
\end{displaymath}
Two limiting cases are familiar.

(i) With equal quark masses
$m_b=m_c\equiv m$, the level shifts become
\begin{eqnarray}
	\Delta^{(2)} & = & \frac{1}{2m^2}\tilde{T}_1 + \frac{1}{m^2}\tilde{T}_2
	-\frac{2}{5m^2}\tilde{T}_4
	\label{shift2} \\
	\Delta^{(0)} & = & -\frac{1}{m^2}\tilde{T}_1 -\frac{2}{m^2}\tilde{T}_2
	-\frac{4}{m^2}\tilde{T}_4
	\nonumber\;\;,
\end{eqnarray}
while the mixing matrix becomes
\begin{equation}
\Delta^{(1)} = \left(\begin{array}{cc}1 & \sqrt{2} \\ \sqrt{2} & 2
\end{array}\right)\left(\frac{-\tilde{T}_1 - 2\tilde{T}_2 +
4\tilde{T}_4}{6m^2} \right) \;\;\; .
\end{equation}
The mass eigenstates are the familiar $^1$P$_1$ and $^3$P$_1$ states of
the $LS$ coupling scheme.  In this basis, they may be written as
\begin{eqnarray}
|^1\text{P}_1\rangle & = &
-{\textstyle{\sqrt{\frac{2}{3}}}}|J_c=\case{3}{2}\rangle
+ {\textstyle{\sqrt{\frac{1}{3}}}}|J_c=\case{1}{2}\rangle \\
|^3\text{P}_1\rangle & = &
{\textstyle{\sqrt{\frac{1}{3}}}}|J_c=\case{3}{2}\rangle
+ {\textstyle{\sqrt{\frac{2}{3}}}}|J_c=\case{1}{2}\rangle \nonumber
\end{eqnarray}
with eigenvalues
\begin{equation}
\left(\begin{array}{c}\lambda(^1$P$_1)\\\lambda(^3$P$_1)
\end{array}\right) = \left(\begin{array}{c}0\\3\end{array}\right)
\left(\frac{-\tilde{T}_1 - 2\tilde{T}_2 + 4\tilde{T}_4}{6m^2} \right)
\;\;\; .
\end{equation}
The position of the $^1\text{P}_1$ level coincides with the centroid
$[5\Delta^{(2)}+3\lambda(^3\text{P}_1)+\Delta^{(0)}]/9$ of the
$^3\text{P}_J$ levels.

(ii) In the heavy-quark limit, $m_b\rightarrow\infty$, the level shifts
of the $J=0,2$ levels become
\begin{eqnarray}
	\Delta^{(2)} & = & \frac{1}{4m_c^2}\tilde{T}_1	\label{shift3} \\
	\Delta^{(0)} & = & -\frac{1}{2m_c^2}\tilde{T}_1 \nonumber\;\;,
\end{eqnarray}
while the mixing matrix becomes
\begin{equation}
\Delta^{(1)} = \left(\!\!\begin{array}{cc}\;1 & \;0 \\ 0 & -2
\end{array}\right)\left(\frac{\tilde{T}_1}{4m_c^2} \right) \;\;\; .
\end{equation}
The $J_c=\frac{3}{2}$ and $J_c=\frac{1}{2}$ states separate into
degenerate pairs, as
expected on the basis of heavy-quark symmetry \cite{hqs}.

In the $c\overline{b}$ system, we label the mass eigenstates obtained by
diagonalizing the matrix (\ref{unequal}) as $n(1^{+})$ and
$n(1^{+\prime})$.  For the 2P$_1$ levels, the mixing matrix is
\begin{equation}
\Delta^{(2\text{P})} = \left(\begin{array}{cc}-1.85\; & \;-2.80 \\ -2.80\;
& \;-4.23 \end{array}\right)\;\;\text{MeV} \;\;\; ,
\end{equation}
with eigenvectors
\begin{eqnarray}
|2(1^+)\rangle & = & 0.552 |J_c=\case{3}{2}\rangle + 0.833
|J_c=\case{1}{2}\rangle \\
|2(1^{+\prime})\rangle & = & -0.833 |J_c=\case{3}{2}\rangle + 0.552
|J_c=\case{1}{2}\rangle \nonumber
\end{eqnarray}
and eigenvalues
\begin{eqnarray}
\lambda_2 & = & -6.09\text{ MeV} \\
\lambda_2^\prime & = & 0.00057\text{ MeV} \nonumber \;\;\; .
\end{eqnarray}
For the 3P$_1$ levels, the mixing matrix is
\begin{equation}
\Delta^{(3\text{P})} = \left(\begin{array}{cc}-0.13\; & \;-2.54 \\ -2.54\;
& \;-6.91 \end{array}\right)\;\;\text{MeV} \;\;\; ,
\end{equation}
with eigenvectors
\begin{eqnarray}
|3(1^+)\rangle & = & 0.316 |J_c=\case{3}{2}\rangle + 0.949
|J_c=\case{1}{2}\rangle \\
|3(1^{+\prime})\rangle & = & -0.949 |J_c=\case{3}{2}\rangle + 0.316
|J_c=\case{1}{2}\rangle \nonumber
\end{eqnarray}
and eigenvalues
\begin{eqnarray}
\lambda_3 & = & -7.76\text{ MeV} \\
\lambda_3^\prime & = & 0.711\text{ MeV} \nonumber \;\;\; .
\end{eqnarray}

For the 4P$_1$ levels, the mixing matrix is
\begin{equation}
\Delta^{(4\text{P})} = \left(\begin{array}{cc}0.71\; & \;-2.44 \\ -2.44\; &
\;-8.31 \end{array}\right)\;\;\text{MeV} \;\;\; ,
\end{equation}
with eigenvectors
\begin{eqnarray}
|4(1^+)\rangle & = & 0.245 |J_c=\case{3}{2}\rangle + 0.969
|J_c=\case{1}{2}\rangle \\
|4(1^{+\prime})\rangle & = & -0.969 |J_c=\case{3}{2}\rangle + 0.245
|J_c=\case{1}{2}\rangle \nonumber
\end{eqnarray}
and eigenvalues
\begin{eqnarray}
\lambda_4 & = & -8.93\text{ MeV} \\
\lambda_4^\prime & = & 1.32\text{ MeV} \nonumber \;\;\; .
\end{eqnarray}

The calculated spectrum of $c\overline{b}$ states is presented in Table
\ref{bcbar} and Figure  \ref{spectre}.  Our spectrum is similar to others
calculated by
Eichten and Feinberg \cite{EF2} in the Cornell potential \cite{Cornell},
by Gershte\u{\i}n et al.
\cite{Gersh2} in the power-law potential (\ref{Martin}), and by Chen and
Kuang \cite{kuangchen} in their own version of a QCD-inspired
potential.  Levels that lie below the $BD$ flavor threshold, i.e., with
$M<M_D+M_B=7.1431\pm0.0021\text{ GeV}\!/\!c^2$, will be stable against
fission into heavy-light mesons.
\subsection{Properties of $c\overline{b}$ Wave Functions at the Origin}
For quarks bound in a central potential, it is convenient to separate
the Schr\"{o}dinger wave function into
radial and angular pieces, as
\begin{equation}
	\Psi_{n\ell m}(\vec{r}) = R_{n\ell}(r)Y_{\ell m}(\theta,\phi)\;\;,
	\label{harm}
\end{equation}
where $n$ is the principal quantum number, $\ell$ and $m$ are the
orbital angular momentum and its projection, $R_{n\ell}(r)$ is the radial
wave function, and $Y_{\ell m}(\theta,\phi)$ is a spherical harmonic
\cite{ylm}.  The Schr\"{o}dinger wave function is normalized,
\begin{equation}
	\int{ d^3\vec{r} |\Psi_{n\ell m}(\vec{r})|^2} = 1 \;\;,
	\label{norm}
\end{equation}
so that
\begin{equation}
	\int_0^\infty r^2 dr |R_{n\ell}(r)| = 1 \;\;.
	\label{norm2}
\end{equation}
The value of the radial wave
function, or its first nonvanishing derivative at the origin,
\begin{equation}
	R_{n\ell}^{(\ell)}(0)\equiv \left.\frac{d^{\,\ell}R_{n\ell}(r)}
	{dr^\ell}\right|_{r=0}\;\;\;,
	\label{wvfc}
\end{equation}
is required to evaluate pseudoscalar decay constants and production rates
through heavy-quark fragmentation \cite{Braaten}.  The quantity
$|R_{n\ell}^{(\ell)}(0)|^2$ is presented for four potentials in Table
\ref{psizero}.  The stronger singularity of the Cornell potential is
reflected in spatially smaller states.

The pseudoscalar decay constant
$f_{B_c}$, which will be required for the discussion of annihilation
decays $c\bar{b}\rightarrow W^+ \rightarrow \text{ final state}$, is defined by
\begin{equation}
\langle 0|A_\mu(0)|B_c(q)\rangle = i f_{B_c} V_{cb} q_\mu\;\;\; ,
\end{equation}
where $A_\mu$ is the axial-vector part of the charged weak current,
$V_{cb}$ is an element of the Cabibbo-Kobayashi-Maskawa quark-mixing
matrix, and $q_\mu$ is the four-momentum of the $B_c$.  The pseudoscalar
decay constant is
related to the ground-state $c\overline{b}$ wave function at the origin by
the van Royen-Weisskopf formula \cite{vrw} modified for color,
\begin{equation}
f_{B_c}^2 = \frac{12|\Psi_{100}(0)|^2}{M} = \frac{3|R_{10}(0)|^2}{\pi M}\;\;\;
{}.
\end{equation}

In the nonrelativistic potential
models we have considered to estimate $M_{B_c}$ and $M_{B_c^*}$, we find
\begin{equation}
f_{B_c} = \left\{\begin{array}{l} 500 \text{ MeV (Buchm\"{u}ller-Tye
potential \cite{BT})} \\ 512 \text{ MeV (power-law potential
\cite{Martin})} \\ 479 \text{ MeV (logarithmic potential \cite{Log})} \\
687 \text{ MeV (Cornell potential \cite{Cornell}).}\end{array}\right.
\end{equation}
Even after QCD radiative corrections of the size suggested by the
comparison of computed and observed leptonic widths for $J\!/\!\psi$ and
$\Upsilon$, $f_{B_c}$ will be significantly larger than the pion decay
constant, $f_\pi=131.74\pm0.15\text{ MeV}$ \cite{PDG}.  The compact size of
the $c\bar{b}$ system enhances the importance of annihilation decays.
\section{Transitions between \lowercase{$c\overline{b}$} States}
\label{sec: Trans}
As in atomic physics, it is the spectral lines produced in cascades from
excited states to the readily observable $B_c$ ground state that will
reveal the $c\bar{b}$ level scheme.  As in the $J\!/\!\psi$ and
$\Upsilon$ quarkonium families, the transitions are mostly radiative
decays.  A few hadronic cascades, analogs of the $2^3\text{S}_1
\rightarrow 1^3\text{S}_1 \pi\pi$ transition first observed in
charmonium, will also be observable.
\subsection{Electromagnetic Transitions}
\label{sec: EM}
Except for the magnetic-dipole (spin-flip) transition between the
ground-state $B_c^*$ and $B_c$, only the electric dipole transitions are
important for mapping the $c\bar{b}$ spectrum.
\subsubsection{Electric Dipole Transitions}
The strength of the electric-dipole transitions is governed by the size
of the radiator and the charges of the constituent quarks.  The E1
transition rate is given by
\begin{equation}
\Gamma_{\text{E1}}(i\rightarrow f+\gamma) = \frac{4\alpha <\!e_Q\!>^2}{27}
k^3 (2J_f+1)|\langle f|r|i\rangle |^2 {\cal S}_{if}\;\;\; ,
\end{equation}
where the mean charge is
\begin{equation}
<\!e_Q\!> = \frac{m_be_c-m_ce_{\overline{b}}}{m_b+m_c} \;\;\; ,
\end{equation}
$k$ is the photon energy, and the statistical factor ${\cal S}_{if}={\cal
S}_{fi}$ is as defined by Eichten and Gottfried \cite{Sif}.  ${\cal
S}_{if}=1$ for $^3$S$_1\rightarrow^3$P$_J$ transitions and ${\cal
S}_{if}=3$ for allowed E1 transitions between spin-singlet states.  The
statistical factors for $d$-wave to $p$-wave transitions are reproduced in
Table \ref{stat} for convenience.
The E1 transition rates and photon energies in the $c\bar{b}$ system are
presented in Table \ref{E1rates}.
\subsubsection{Magnetic Dipole Transitions}
The only decay mode for the 1$^3$S$_1$ $(B_c^*)$ state is the magnetic
dipole transition to the ground state, $B_c$.  The M1 rate for transitions
between $s$-wave levels is given by
\begin{equation}
\Gamma_{\text{M1}}(i\rightarrow f+\gamma) = \frac{16\alpha}{3}\mu^2 k^3
(2J_f+1)|\langle f|j_0(kr/2)|i\rangle|^2\;\;\;,
\end{equation}
where the magnetic dipole moment is
\begin{equation}
\mu = \frac{m_be_c-m_ce_{\overline{b}}}{4m_cm_b}\;\;\;
\end{equation}
and $k$ is the photon energy.  Rates for the allowed and hindered M1
transitions between spin-triplet and spin-singlet $s$-wave $c\overline{b}$
states are given in Table \ref{M1rates}.  The M1 transitions contribute
little to the total widths of the 2S levels.  Because it cannot decay by
annihilation, the 1$^3$S$_1$ $c\overline{b}$ level, with a total width of
135~eV, is far more stable than its counterparts in the $c\overline{c}$
and $b\overline{b}$ systems, whose total widths are $68 \pm 10$~keV and
$52.1 \pm 2.1$~keV, respectively \cite{PDG}.
\subsection{Hadronic Transitions}
A hadronic transition between quarkonium levels can be understood as a
two-step process in which gluons first are emitted from the heavy quarks
and then recombine into light hadrons.  Perturbative QCD is not directly
applicable, because the energy available to the light hadrons is small and
the emitted gluons are soft.  Nevertheless, the final quarkonium state is
small compared to the system of light hadrons and moves nonrelativistically
in the rest frame of the decaying quarkonium state.  A multipole expansion
of the color gauge field converges rapidly and leads to selection rules, a
Wigner-Eckart theorem, and rate estimates for hadronic transitions
\cite{Gott78}.  The recombination of gluons into light hadrons involves the
full strong dynamics and can only be modeled.  The general structure of
hadronic-cascade transitions and models for the recombination of gluons
into light hadrons can be found in a series of papers by Yan and
collaborators \cite{Yan80,Yan_bb,Yan_1P1,Yan_2D}.

The hadronic transition rates for an unequal-mass $Q\bar{Q}^\prime$ system
differ in some details from the rates for an equal-mass $Q\bar{Q}$ system with
the same reduced mass.
The relative strengths of various terms that contribute to
magnetic-multipole transitions are modified because of the unequal quark
and antiquark masses.  The electric-multipole transitions
are only sensitive to the relative position of the
quark and antiquark and will be unchanged in form.

As in the $c\bar{c}$ and $b\bar{b}$ systems, the principal hadronic
transitions in the $c\bar{b}$ system involve the emission of two pions.
Electric-dipole contributions dominate in these transitions, and so the
equal-mass results apply directly.  The initial quarkonium state is
characterized by its total angular momentum $J^\prime$ with $z$-component
$M^\prime$, orbital angular momentum $\ell^\prime$, spin $s^\prime$, and
other quantum numbers collectively labelled by $\alpha^\prime$.  The
corresponding quantum numbers of the final quarkonium state are denoted by
the unprimed symbols.  Since the transition operator is spin-independent,
the initial and final spins are the same: $s^\prime=s$.  Because the
gauge-field operators in the transition amplitude do not depend on the
heavy-quark variables, the transition operator is a reducible second-rank
tensor, which may be decomposed into a sum of irreducible tensors with rank
$k=0,1,2$.  The differential rate \cite{Yan80} for the E1--E1 transition
from the initial quarkonium state $\Phi^\prime$ to the final quarkonium
state $\Phi$ and a system of $n$ light hadrons, denoted $h$, is given by
\begin{equation}
 \frac{d\Gamma}{d{\cal M}^2}(\Phi^\prime \rightarrow \Phi + h) =
    (2J+1)\sum_{k=0}^2 \left\{ {\begin{array}{rrr} k & \ell^\prime& \ell \\
    s & J & J' \end{array}} \right\}^2 A_k(\ell^\prime,\ell) \; ,
\label{WET}
\end{equation}
where ${\cal M}^2$ is the invariant mass squared of the light hadron
system, $\left\{\;\right\}$ is a 6-$j$ symbol, and $A_k(\ell^\prime,\ell)$
is the contribution of the irreducible tensor with rank $k$.
The Wigner-Eckart theorem (\ref{WET}) yields the relations among two-pion
transition rates  given in Table \ref{hadI}.

The magnitudes of the $A_k(\ell^\prime,\ell)$ are model-dependent.  Since
the $A_1$ contributions are suppressed in the soft-pion limit \cite{Yan80},
we will set $A_1(\ell^\prime,\ell) = 0$.  For some of the remaining rates
we can use simple scaling arguments from the measured rates in $Q\bar Q$
systems \cite{detail}.  The amplitude for an E1--E1 transition depends
quadratically on the interquark separation, so the scaling law between a
$Q\bar{Q}^\prime$ and the corresponding $Q\bar Q$ system states is given by
\cite{Gott78,Yan80}:
\begin{equation}
 \frac{\Gamma (Q\bar{Q}^\prime)}{\Gamma (Q\bar{Q})}
   = \frac{\langle r^2(Q\bar{Q}^\prime)\rangle^2}{\langle
r^2(Q\bar{Q})\rangle^2}\;\;,
\end{equation}
up to possible differences in phase space.  The measured values for the
$\psi^\prime \rightarrow \psi + \pi\pi$, $\Upsilon^\prime \rightarrow
\Upsilon + \pi\pi$, and $\psi (3770) \rightarrow \psi + \pi\pi$ transition
rates allow good scaling estimates for the $2\text{S}\rightarrow 1\text{S}
+ \pi\pi$ and $3\text{D}\rightarrow 1\text{S} + \pi\pi$ transitions in the
$c\bar{b}$ system.  We have estimated the remaining transition rates by
scaling the $b \bar b$ rates calculated by Kuang and Yan \cite{Yan_bb} in
their Model C, which is based on the Buchm\"{u}ller-Tye potential
\cite{BT}.  The results are shown in Table \ref{hadII}.

Chiral symmetry leads to a universal form for the normalized dipion
spectrum \cite{BC},
\begin{equation}
  \frac{1}{\Gamma} \frac{d\Gamma}{d{\cal{M}}} = \text{Constant}\times
  \frac{|\vec{K}|}{M_{\Phi^\prime}^2}(2x^2-1)^2\sqrt{x^2-1} \;\;,
          \label{BandC}
\end{equation}
where $x = {\cal{M}}/2m_{\pi}$ and
\begin{equation}
|\vec{K}| = \frac{\sqrt{M_{\Phi^\prime}^2 - ({\cal M}+M_\Phi)^2}
          \sqrt{M_{\Phi^\prime}^2 - ({\cal M}-M_\Phi)^2}}{2M_{\Phi^\prime}}
\end{equation}
is the three-momentum carried by the pion pair.  The normalized
invariant-mass distribution for the transition $2^3\text{S}_1\rightarrow
1^3\text{S}_1+\pi\pi$ is shown in Figure \ref{hadfig} for the $c\bar{c}$,
$c\bar{b}$, and $b\bar{b}$ families.  The soft-pion expression
(\ref{BandC}) describes the depletion of the dipion spectrum
at low invariant masses observed in the transitions
$\psi(2\text{S})\rightarrow \psi(1\text{S})\pi\pi$ \cite{pipidk} and
$\Upsilon(2\text{S})\rightarrow \Upsilon(1\text{S})\pi\pi$
\cite{pipidk1}, but fails to
account for the $\Upsilon(3\text{S})\rightarrow
\Upsilon(1\text{S})\pi\pi$ and $\Upsilon(3\text{S})\rightarrow
\Upsilon(2\text{S})\pi\pi$ spectra \cite{pipidk2}.  We expect the 3S
levels to lie above flavor threshold in the $c\bar{b}$ system.

By the Wigner-Eckart theorem embodied in Eq. (\ref{WET}), the
invariant mass spectrum in the decay $B_c\text{(2S)}\rightarrow
B_c\text{(1S)} +\pi\pi$ should have the same form (\ref{BandC}) as the
$B_c^*\text{(2S)}\rightarrow B_c^*\text{(1S)} +\pi\pi$ transition.
Braaten, Cheung, and Yuan \cite{Braaten} have calculated the probability
for a high-energy $\bar{b}$ antiquark to fragment into the $c\bar{b}$
$s$-waves as $3.8\times 10^{-4}\text{ for } \bar{b}\rightarrow
B_c\text{(1S)}$, $5.4\times 10^{-4}\text{ for } \bar{b}\rightarrow
B_c^*\text{(1S)}$, $2.3\times 10^{-4}\text{ for } \bar{b}\rightarrow
B_c\text{(2S)}$, and $3.2\times 10^{-4}\text{ for } \bar{b}\rightarrow
B_c^*\text{(2S)}$.  Given the excellent experimental signatures for
$B_c\text{(1S)}$ decay and the favorable prospects for $B_c\text{(2S)}$
production in high-energy proton-antiproton collisions, it may be possible
to observe the $0\rightarrow 0$ transition for the first time in the $B_c$
family.

The 2$^3$S$_1\rightarrow$ 1$^3$S$_1 + \eta$ transition
has been observed in charmonium.  This transition proceeds
via an M1--M1 or E1--M2 multipole.
In the $c\bar{b}$ system the E1--M2 multipole dominates and
the scaling from the $c\bar{c}$ system should be given by
\begin{equation}
 \frac{\Gamma (c\bar{b})}{\Gamma (c\bar{c})} =
\frac{(m_b+m_c)^2}{4m_b^2}\frac{\langle r^2(c\bar{b})\rangle}{
\langle r^2(c\bar c)\rangle} \frac{M_{\psi^\prime}^3}{M_{\Phi^\prime}^3}
\frac{[M_{\Phi^\prime}^2-(M_\Phi+M_\eta)^2]^{1/2}
[M_{\Phi^\prime}^2-(M_\Phi-M_\eta)^2]^{1/2}}
{[M_{\psi^\prime}^2-(M_\psi+M_\eta)^2]^{1/2}
[M_{\psi^\prime}^2-(M_\psi-M_\eta)^2]^{1/2}}\;\;\; ,
\end{equation}
where $M_{\Phi^\prime}$ and $M_\Phi$ are the masses of the 2$^3$S$_1$ and
1$^3$S$_1$
$c\bar{b}$ levels, respectively.  Because of the small energy release in
this transition, the slightly smaller level spacing in the $B_c$ family
compared to the $J\!/\!\psi$ family (562 MeV {\it vs.\/} 589 MeV)
strongly suppresses $\eta$-emission in the $c\bar{b}$ system.
The observed rate of $\Gamma (\psi^\prime \rightarrow \psi + \eta)
= 6.6 \pm 2.1\text{ keV}$ \cite{PDG} scales to $\Gamma (B_c\text{(2S)}
\rightarrow B_c\text{(1S)} + \eta) =0.25\text{ keV}$.

\subsection{Total Widths and Experimental Signatures}
The total widths and branching fractions are given in Table
\ref{totalwidth}.  The most striking feature of the $c\bar{b}$ spectrum
is the extreme narrowness of the states.  A crucial element in unraveling
the spectrum will be the efficient detection of the 72-MeV M1-photon
that, in coincidence with an observed $B_c$ decay, tags the $B_c^*$.
This will be essential for distinguishing the
$B_c\text{(2S)}\rightarrow B_c\text{(1S)} +\pi\pi$ transition from
$B_c^*\text{(2S)}\rightarrow B_c^*\text{(1S)} +\pi\pi$, which will have
a nearly identical spectrum and a comparable rate.  Combining the branching
fractions in Table \ref{totalwidth} with the $b$-quark fragmentation
probabilities of Ref. \cite{Braaten}, we expect the cross section times
branching fractions to be in the proportions
\begin{equation}
	\sigma B(B_c\text{(2S)}\rightarrow B_c\text{(1S)} +\pi\pi) \approx
	1.2\times \sigma B(B_c^*\text{(2S)}\rightarrow B_c^*\text{(1S)}
	+\pi\pi)\;\;.
	\label{sigmaB}
\end{equation}

A reasonable---but challenging---experimental goal would be to map the
eight lowest-lying $c\bar{b}$ states: the 1S, 2S, and 2P levels.  A first
step, in addition to reconstructing the hadronic cascades we have just
discussed, would be the detection of the 455-MeV photons in coincidence
with $B_c$, and of 353-, 382-, and 397-MeV photons in coincidence with
$B_c^*\rightarrow B_c+\gamma(72\text{ MeV})$.  This would be a most
impressive triumph of experimental art.

\section{Concluding Remarks}
A meson with beauty and charm is an exotic particle, but prospects are
good that it will be discovered in the near future.  As soon as $B_c$ has
been identified, the investigation of competing weak-decay mechanisms,
$\bar{b}\rightarrow\bar{c}W^+$ (represented by $\psi\pi^+$,
$\psi\ell^+\nu$, etc.), $c\rightarrow sW^+$ (represented by $B_s\pi^+$,
$B_s\ell^+\nu$, etc.), and $c\bar{b}\rightarrow W^+$ (represented by
$\psi D_s^+$, $\tau^+\nu_\tau$, etc.), can begin.  The issues to be studied,
and predictions for a wide variety of inclusive and exclusive decays, are
presented in a companion paper \cite{eqweak}.  Before the end of the
decade, it should prove possible to map out part of the $c\bar{b}$
spectrum by observing $\gamma$- and $\pi\pi$-coincidences with the
ground-state $B_c$ or its hyperfine partner $B_c^*$.
\acknowledgments
Fermilab is operated by Universities Research Association, Inc., under
contract DE-AC02-76CHO3000 with the United States Department of Energy.
C.Q. thanks the Cultural Section of the Vienna municipal government and
members of the Institute for Theoretical Physics of the
University of Vienna for their warm hospitality while part of this work
was carried out.

\begin{table}
\caption{Quarkonium ground-state masses (in $\text{GeV}\!/\!c^2$) in three
potentials.}
\begin{tabular}{ccccc}
Observable & QCD, Ref. \cite{BT} & Power-law, Ref. \cite{Martin} &
Logarithmic, Ref. \cite{Log} &  Cornell, Ref. \cite{Cornell} \\
\tableline
$(c\bar{c})$ 1S & 3.067 & 3.067 & 3.067 & 3.067\\
$\psi$ & 3.097 & 3.097 & 3.097 & 3.097 \\
$\eta_c$ & 2.980 & 2.980 & 2.980 & 2.980 \\
$\psi - \eta_c$ & 0.117\tablenotemark[1] & 0.117\tablenotemark[2] &
0.117\tablenotemark[3] & 0.117\tablenotemark[4] \\
\hline
$(c\bar{b})$  1S & 6.317 & 6.301 & 6.317 & 6.321 \\
$B_c^*$ & 6.337 & 6.319 & 6.334 & 6.343 \\
$B_c$ & 6.264 & 6.248 & 6.266 & 6.254 \\
$B_c^*-B_c$ & 0.073 & 0.071 & 0.068 & 0.089 \\
\hline
$(b\bar{b})$  1S & 9.440 & 9.446 & 9.444 & 9.441 \\
$\Upsilon$ & 9.464 & 9.462 & 9.460 & 9.476 \\
$\eta_b$ & 9.377 & 9.398 & 9.395 & 9.335 \\
$\Upsilon - \eta_b$ & 0.087 & 0.064 & 0.065 & 0.141
\end{tabular}
\label{Mass1S}
\tablenotemark[1]{Input value; determines $\alpha_s=0.36$.}
\tablenotemark[2]{Input value; determines $\alpha_s=0.43$.}
\tablenotemark[3]{Input value; determines $\alpha_s=0.37$.}
\tablenotemark[4]{Input value; determines $\alpha_s=0.31$.}
\end{table}

\begin{table}
\caption{Charmonium masses and leptonic widths in the Buchm\"{u}ller-Tye
potential.}
\begin{tabular}{ccccc}
Level & \multicolumn{2}{c}{Mass ($\text{GeV}\!/\!c^2$)} &
\multicolumn{2}{c}{Leptonic Width (keV)} \\
 & Calculated & Observed\tablenotemark[1] & Calculated &
 Observed\tablenotemark[1]\\
\tableline
1$^1$S$_0$ $(\eta_c)$ & 2.980 & $2.9788 \pm 0.0019$ & & \\
1$^3$S$_1$ $(\psi/J)$& 3.097 & $3.09688 \pm 0.00001 \pm
0.00006$\tablenotemark[2] &
8.00 & $4.72 \pm 0.35$ \\
2$^3$P$_0$ $(\chi_{c0})$& 3.436 & $3.4151  \pm  0.0010$ & & \\
2$^3$P$_1$ $(\chi_{c1})$& 3.486 & $3.51053  \pm  0.00004\pm
0.00012$\tablenotemark[2] & & \\
2$^3$P$_2$ $(\chi_{c2})$& 3.507 & $3.55615  \pm 0.00007 \pm
0.00012$\tablenotemark[2]  & &\\
2$^1$P$_1$ $(h_c)$& 3.493 & $3.5262  \pm 0.00015\pm0.0002$\tablenotemark[3]  &
& \\
2$^1$S$_0$ $(\eta_c^\prime)$& 3.608 &  & & \\
2$^3$S$_1$ $(\psi^\prime)$& 3.686 & $3.68600  \pm 0.00010 $ & 3.67 & $2.14
\pm 0.21$
\end{tabular}
\label{charmonium}
\tablenotemark[1]{See Ref. \cite{PDG}.}
\tablenotemark[2]{See Ref. \cite{chistates}.}
\tablenotemark[3]{See Ref. \cite{Rosanna}.}
\end{table}

\begin{table}
\caption{$b\bar{b}$ masses and leptonic widths in the
Buchm\"{u}ller-Tye potential.}
\begin{tabular}{ccccc}
Level & \multicolumn{2}{c}{Mass ($\text{GeV}\!/\!c^2$)} &
\multicolumn{2}{c}{Leptonic Width (keV)} \\
 & Calculated & Observed\tablenotemark[1] & Calculated &
 Observed\tablenotemark[1]\\
\tableline
1$^1$S$_0$ $(\eta_b)$ & 9.377 & & &\\
1$^3$S$_1$ $(\Upsilon)$& 9.464 & $9.46032 \pm 0.00022$ & 1.71 & $1.34 \pm
0.04$ \\
2$^3$P$_0$ $(\chi_{b0})$& 9.834 & $9.8598  \pm  0.0013$  & & \\
2$^3$P$_1$ $(\chi_{b1})$& 9.864 & $9.8919  \pm  0.0007$  & & \\
2$^3$P$_2$ $(\chi_{b2})$& 9.886 & $9.9132  \pm  0.0006$  & & \\
2$^1$P$_1$ $(h_b)$& 9.873 &  & & \\
2$^1$S$_0$ $(\eta_b^\prime)$& 9.963 &  & & \\
2$^3$S$_1$ $(\Upsilon^\prime)$& 10.007 & $10.02330  \pm 0.00031 $ & 0.76 &
$0.586 \pm 0.029$ \\
3$^3$D$_1$ & 10.120 & & &  \\
3$^3$D$_2$ & 10.126 & & &  \\
3$^3$D$_3$ & 10.130 & & &  \\
3$^1$D$_2$ & 10.127 & & &  \\
3$^3$P$_0$ $(\chi_{b0})$& 10.199 & $10.2320  \pm  0.0007$ & & \\
3$^3$P$_1$ $(\chi_{b1})$& 10.224 & $10.2549  \pm  0.0006$ & & \\
3$^3$P$_2$ $(\chi_{b2})$& 10.242 & $10.26835  \pm  0.00057$ & & \\
3$^1$P$_1$ $(h_b)$& 10.231 & & &  \\
3$^1$S$_0$ & 10.298 & & &  \\
3$^3$S$_1$ & 10.339 & $10.3553  \pm 0.0005$ & 0.55 & $0.44 \pm 0.03$ \\
4$^1$S$_0$ & 10.573 &  & & \\
4$^3$S$_1$ & 10.602 & $10.5800  \pm 0.0035$ & &
\end{tabular}
\tablenotemark[1]{See Ref. \cite{PDG}.}
\label{upsilons}
\end{table}

\begin{table}
\caption{$c\bar{b}$ masses (in $\text{GeV}\!/\!c^2$) in the Buchm\"{u}ller-Tye
potential.}
\begin{tabular}{ccccc}
Level & Calculated Mass & Eichten \& Feinberg\tablenotemark[1] &
Gershte\u{\i}n et al.\tablenotemark[2] & Chen \& Kuang\tablenotemark[3] \\
\tableline
1$^1$S$_0$ $(B_c)$ & 6.264 & 6.243 & 6.246 & 6.310\\
1$^3$S$_1$ $(B_c^*)$ & 6.337 & 6.339 & 6.329 & 6.355 \\
2$^3$P$_0$ & 6.700 & 6.697 & 6.645 & 6.728  \\
2 $1^{+\prime}$ & 6.736 & 6.740 & 6.741 & 6.760 \\
2 $1^{+}$ & 6.730  & 6.719 & 6.682 & 6.764 \\
2$^3$P$_2$ & 6.747  & 6.750 & 6.760 & 6.773 \\
2$^1$S$_0$ & 6.856  & 6.969 & 6.863 & 6.890 \\
2$^3$S$_1$ & 6.899  & 7.022 & 6.903 & 6.917 \\
3$^3$D$_1$ & 7.012  &  &  & \\
3$^3$D$_2$ & 7.012  &  &  & \\
3$^3$D$_3$ & 7.005  &  & (7.008) & \\
3$^1$D$_2$ & 7.009  &  &  & \\
3$^3$P$_0$ & 7.108  &  & 7.067 & 7.134 \\
3 $1^{+\prime}$ & 7.142 &  & 7.129 & 7.159 \\
3 $1^{+}$ & 7.135 &  & 7.099 & 7.160 \\
3$^3$P$_2$ & 7.153 &  & 7.143 & 7.166 \\
3$^1$S$_0$ & 7.244 &  & (7.327) &  \\
3$^3$S$_1$ & 7.280 &  &  &  \\
4$^1$S$_0$ & 7.562 &  &  &  \\
4$^3$S$_1$ & 7.594 &  &  &
\end{tabular}
\label{bcbar}
\tablenotemark[1]{See Ref. \cite{EF2}.}
\tablenotemark[2]{See Ref. \cite{Gersh2}.}
\tablenotemark[3]{See Ref. \cite{kuangchen}; the masses
correspond to Potential I with $\Lambda_{\overline{\text{MS}}} =
150\text{ MeV}$.}
\end{table}

\begin{table}
\caption{Radial wave functions at the origin and related quantities for
$c\bar{b}$ mesons.}
\begin{tabular}{ccccc}
Level & \multicolumn{3}{c}{$|R_{n\ell}^{(\ell)}(0)|^2$} \\
 & QCD, Ref. \cite{BT} & Power-law, Ref. \cite{Martin} & Logarithmic,
 Ref. \cite{Log} & Cornell, Ref. \cite{Cornell}  \\
\tableline
1S & $1.642\text{ GeV}^3$ & $1.710\text{ GeV}^3$ & $1.508\text{ GeV}^3$ &
$3.102\text{ GeV}^3$ \\
2P & $0.201\text{ GeV}^5$ & $0.327\text{ GeV}^5$ & $0.239\text{ GeV}^5$ &
$0.392\text{ GeV}^5$ \\
2S & $0.983\text{ GeV}^3$ & $0.950\text{ GeV}^3$ & $0.770\text{ GeV}^3$ &
$1.737\text{ GeV}^3$ \\
3D & $0.055\text{ GeV}^7$ & $0.101\text{ GeV}^7$ & $0.055\text{ GeV}^7$ &
$0.080\text{ GeV}^7$ \\
3P & $0.264\text{ GeV}^5$ & $0.352\text{ GeV}^5$ & $0.239\text{ GeV}^5$ &
$0.531\text{ GeV}^5$ \\
3S & $0.817\text{ GeV}^3$ & $0.680\text{ GeV}^3$ & $0.563\text{ GeV}^3$ &
$1.427\text{ GeV}^3$
\end{tabular}
\label{psizero}
\end{table}

\begin{table}
\caption{Statistical Factor ${\cal S}_{if}$ for
$^3$P$_J\rightarrow^3$D$_{J^\prime}+\gamma$ Transitions.}
\begin{tabular}{ccc}
$J$ & $J^\prime$ & ${\cal S}_{if}$ \\
\tableline
0 & 1 & 2 \\
1 & 1 & 1/2 \\
1 & 2 & 9/10 \\
2 & 1 & 1/50 \\
2 & 2 & 9/50 \\
2 & 3 & 18/25
\end{tabular}
\label{stat}
\end{table}

\begin{table}
\renewcommand{\baselinestretch}{1.0}
\caption{E1 Transition Rates in the $c\bar{b}$ System.}
\begin{tabular}{ccdd}

Transition & Photon energy (MeV) & $\langle f|r|i\rangle~({\rm GeV}^{-1})$
& $\Gamma(i\rightarrow f+\gamma)$~(keV) \\
\tableline
2$^3$P$_2 \rightarrow$ 1$^3$S$_1+\gamma$  & 397  & 1.714 &
112.6  \\
2$(1^{+})\rightarrow$ 1$^3$S$_1+\gamma$  & 382  & 1.714  &99.5  \\
2$(1^{+})\rightarrow$  1$^1$S$_0+\gamma$  & 450  & 1.714 &0.0 \\
2$(1^{+\prime})\rightarrow$ 1$^3$S$_1+\gamma$ & 387  & 1.714 &0.1  \\
2$(1^{+\prime})\rightarrow$  1$^1$S$_0+\gamma$  & 455  & 1.714 &56.4
\\
2$^3$P$_0\rightarrow$ 1$^3$S$_1+\gamma$ & 353  & 1.714 &79.2 \\
2$^3$S$_1\rightarrow$ 2$^3$P$_2+\gamma$ & 151  & $-$2.247 &17.7 \\
2$^3$S$_1\rightarrow$ 2$(1^{+})+\gamma$  &167  & $-$2.247 &14.5 \\
2$^3$S$_1\rightarrow$ 2$(1^{+\prime})+\gamma$  & 161  & $-$2.247 &0.0 \\
2$^3$S$_1\rightarrow$ 2$^3$P$_0+\gamma$  & 196  & $-$2.247 &7.8 \\
2$^1$S$_0\rightarrow$  2$(1^{+})+\gamma$  & 125  & $-$2.247 &0.0 \\
2$^1$S$_0\rightarrow$  2$(1^{+\prime})+\gamma$  & 119  & $-$2.247 & 5.2 \\
3$^3$D$_3\rightarrow$ 2$^3$P$_2+\gamma$  & 258  & 2.805 &98.7 \\
3$^3$D$_2\rightarrow$ 2$^3$P$_2+\gamma$ & 258  & 2.805 &24.7 \\
3$^3$D$_2\rightarrow$ 2$(1^{+})+\gamma$  &274  & 2.805 &88.8 \\
3$^3$D$_2\rightarrow$ 2$(1^{+\prime})+\gamma$  & 268  & 2.805 &0.1 \\
3$^3$D$_1\rightarrow$ 2$^3$P$_2+\gamma$  & 258  & 2.805 &2.7 \\
3$^3$D$_1\rightarrow$ 2$(1^{+})+\gamma$  & 274  & 2.805 &49.3 \\
3$^3$D$_1\rightarrow$  2$(1^{+\prime})+\gamma$  & 268  & 2.805 &0.0 \\
3$^3$D$_1\rightarrow$  2$^3$P$_0+\gamma$  & 302  & 2.805 &88.6 \\
3$^1$D$_2\rightarrow$  2$(1^{+\prime})+\gamma$  & 268  & 2.805 &92.5
\\
3$^3$P$_2\rightarrow$  1$^3$S$_1+\gamma$  & 770  & 0.304 &25.8 \\
3$^3$P$_2\rightarrow$  2$^3$S$_1+\gamma$  & 249  & 2.792 &73.8 \\
3$^3$P$_2\rightarrow$  3$^3$D$_3+\gamma$  & 142  & $-$2.455 &17.8 \\
3$^3$P$_2\rightarrow$  3$^3$D$_2+\gamma$  & 142  & $-$2.455 &3.2 \\
3$^3$P$_2\rightarrow$  3$^3$D$_1+\gamma$  & 142  & $-$2.455 &0.2 \\
3$(1^{+})\rightarrow$  1$^3$S$_1+\gamma$  & 754  & 0.304 &22.1 \\
3$(1^{+})\rightarrow$  2$^3$S$_1+\gamma$  & 232  & 2.792 &54.3 \\
3$(1^{+})\rightarrow$  3$^3$D$_2+\gamma$  & 125  & $-$2.455 &9.8 \\
3$(1^{+})\rightarrow$  3$^3$D$_1+\gamma$  & 125  & $-$2.455 &0.3 \\
3$(1^{+\prime})\rightarrow$  1$^3$S$_1+\gamma$  & 760  & 0.304 &2.1 \\
3$(1^{+\prime})\rightarrow$  2$^3$S$_1+\gamma$  &239  & 2.792 &5.4 \\
3$(1^{+\prime})\rightarrow$  3$^3$D$_2+\gamma$  & 131  & $-$2.455 &11.5
\\
3$(1^{+\prime})\rightarrow$  3$^3$D$_1+\gamma$  & 131  & $-$2.455 &0.4
\\
3$^3$P$_0\rightarrow$  1$^3$S$_1+\gamma$  & 729  & 0.304 &21.9 \\
3$^3$P$_0\rightarrow$  2$^3$S$_1+\gamma$  & 205  & 2.792 &41.2 \\
3$^3$P$_0\rightarrow$  3$^3$D$_1+\gamma$  & 98  & $-$2.455 &6.9
\end{tabular}
\label{E1rates}
\end{table}

\begin{table}
\caption{M1 Transition Rates in the $c\bar{b}$ System.}
\begin{tabular}{ccdd}
Transition & Photon energy (MeV) & $\langle f|j_0(kr/2)|i\rangle$ &
$\Gamma(i\rightarrow f+\gamma)$~(keV)\\
\tableline
2$^3$S$_1 \rightarrow$ 2$^1$S$_0+\gamma$ & 43 & 0.9990 & 0.0289\\
2$^3$S$_1 \rightarrow$ 1$^1$S$_0+\gamma$ & 606 & 0.0395 & 0.1234\\
2$^1$S$_0 \rightarrow$ 1$^3$S$_1+\gamma$ & 499 & 0.0265 & 0.0933\\
1$^3$S$_1 \rightarrow$ 1$^1$S$_0+\gamma$ & 72 & 0.9993 & 0.1345
\end{tabular}
\label{M1rates}
\end{table}
\begin{table}
\caption{The relative rates for the allowed two-pion
E1--E1 transitions between spin-triplet states and spin-singlet
states.
The reduced rates are denoted by $A_{k}(\ell^\prime,\ell)$
where $k$ is the rank of the irreducible tensor for gluon emission and
$\ell^\prime$ and $\ell$ are the orbital angular momenta of the initial
and final states respectively.}
\begin{tabular}{ccc}
Transition &  Rate & $c\bar{b}$ Estimate (keV)\tablenotemark[1]\\
\hline
3$^3$P$_2\rightarrow $ 2$^3$P$_2 +\pi\pi$ & $A_0(1,1)/3 + A_1(1,1)/4 + 7
A_2(1,1)/60$ & 1.4 \\
3$^3$P$_2\rightarrow $ 2$^3$P$_1 +\pi\pi$ & $A_1(1,1)/12 + 3 A_2(1,1)/20$
& 0.03 \\
3$^3$P$_2\rightarrow $ 2$^3$P$_0 +\pi\pi$ & $A_2(1,1)/15$ & 0.01\\
3$^3$P$_1\rightarrow $ 2$^3$P$_2 +\pi\pi$ & $5 A_1(1,1)/36 + A_2(1,1)/4$ &
0.05\\
3$^3$P$_1\rightarrow $ 2$^3$P$_1 +\pi\pi$ & $A_0(1,1)/3 + A_1(1,1)/12 +
A_2(1,1)/12$ & 0.02\\
3$^3$P$_1\rightarrow $ 2$^3$P$_0 +\pi\pi$ & $A_1(1,1)/9$ & 0\\
3$^3$P$_0\rightarrow $ 2$^3$P$_2 +\pi\pi$ & $A_2(1,1)/3$ & 0.07\\
3$^3$P$_0\rightarrow $ 2$^3$P$_1 +\pi\pi$ & $A_1(1,1)/3$ & 0\\
3$^3$P$_0\rightarrow $ 2$^3$P$_0 +\pi\pi$ & $A_0(1,1)/3$ & 1.4\\
3$^3$D$_{J^\prime}\rightarrow $ 1$^3$S$_1 +\pi\pi$ & $A_2(2,0)/5$ & $32
\pm 11$\\
2$^3$S$_1\rightarrow $ 1$^3$S$_1 +\pi\pi$ & $A_0(0,0)$ & $50 \pm 7$ \\
\hline
3$^1$P$_1\rightarrow $ 2$^1$P$_1 +\pi\pi$ & $A_0(1,1)/3 + A_1(1,1)/3 +
A_2(1,1)/3$ & 1.4 \\
3$^1$D$_2\rightarrow $ 1$^1$S$_0 +\pi\pi$ & $A_2(2,0)/5$ & $32 \pm 11$\\
2$^1$S$_0\rightarrow $ 1$^1$S$_0 +\pi\pi$ & $A_0(0,0)$ & $50 \pm 7$ \\
\end{tabular}
\tablenotemark[1]{Sum of $\pi^+\pi^-$ and $\pi^0\pi^0$.}
\label{hadI}
\end{table}

\begin{table}
\caption{Estimated rates for two-pion E1--E1 transitions between
$c\bar{b}$ levels, scaled from $c\bar{c}$ and $b\bar{b}$ measurements and
calculations.}
\begin{tabular}{cccc}
Transition & $(Q\bar{Q})$ rate (keV)
& $\langle r^2(c\bar{b})\rangle/\langle r^2(Q\bar{Q})\rangle$ &
Reduced rate ($c\bar{b}$) (keV) \\
\hline
 & $(b\bar{b}):~11.7 \pm 2.2^{\rm a}$ & 1.99 &
 $A_0(0,0) = 40 \pm 8$\\
2$^3$S$_1 \rightarrow $1$^3$S$_1 + \pi\pi $  & $(c\bar{c}):~141 \pm
27$\tablenotemark[1] &  0.70 & $A_0(0,0) = 69 \pm 13$ \\
   & Mean & & $A_0(0,0) = 50\pm 7$ \\
   \hline
 & $(c\bar{c}):~37 \pm 17 \pm 8$\tablenotemark[2] &
   & $A_2(2,0) = 137 \pm 70$\\
3$^3$D$_1 \rightarrow $1$^3$S$_1 + \pi\pi $ & $(c\bar{c}):~55 \pm 23
\pm 11$\tablenotemark[3]  & 0.72 & $A_2(2,0) = 204 \pm 94$ \\
  & Mean:~$43 \pm 15$ & & $A_2(2,0) = 160 \pm 56$ \\
\hline
3$^3$P$_0 \rightarrow$ 2$^3$P$_0 + \pi\pi $ &
$(b\bar{b}):~0.4$\tablenotemark[4]  & 1.88   & $A_0(1,1) = 4.2$ \\
\hline
3$^3$P$_2 \rightarrow$ 2$^3$P$_1 + \pi\pi $ & $(b\bar{b}):~0.01^{\rm
d}$  & 1.88
   & $A_2(1,1) = 0.2$ \\
\end{tabular}
\tablenotemark[1]{Particle Data Group average \cite{PDG}.}
\tablenotemark[2]{Measured by the Crystal Ball \cite{Xball} and Mark II
 \cite{MarkII} Collaborations.}
\tablenotemark[3]{Measured by the Mark III Collaboration \cite{MarkIII}.}
\tablenotemark[4]{Calculated by Kuang and Yan \cite{Yan_bb}
using the Buchm\"uller-Tye potential \cite{BT}. }
\label{hadII}
\end{table}

\begin{table}
\squeezetable
\renewcommand{\baselinestretch}{0.95}
\caption{Total widths and branching fractions of $c\bar{b}$ levels.}

\begin{tabular}{cd}

Decay Mode & Branching Fraction (percent) \\
\tableline
\multicolumn{2}{c}{1$^3$S$_1$: $\Gamma = 0.135\text{ keV}$} \\
1$^1$S$_0+\gamma$ & 100. \\
\tableline
\multicolumn{2}{c}{2$^1$S$_0$: $\Gamma = 55\text{ keV}$} \\
1$^1$S$_0+\pi\pi$ & 91. \\
2$(1^{+\prime})+\gamma$ & 9. \\
\tableline
\multicolumn{2}{c}{2$^3$S$_1$: $\Gamma = 90\text{ keV}$} \\
1$^3$S$_1+\pi\pi$ & 55. \\
2$^3$P$_2+\gamma$ & 20. \\
2$(1^+)+\gamma$ & 16. \\
2$^3$P$_0+\gamma$ & 9. \\
\tableline
\multicolumn{2}{c}{2$^3$P$_0$: $\Gamma = 79\text{ keV}$} \\
1$^3$S$_1+\gamma$ & 100. \\
\tableline
\multicolumn{2}{c}{2$(1^+)$: $\Gamma = 100\text{ keV}$} \\
1$^3$S$_1+\gamma$ & 100. \\
\tableline
\multicolumn{2}{c}{2$(1^{+\prime})$: $\Gamma = 56\text{ keV}$} \\
1$^1$S$_0+\gamma$ & 100. \\
\tableline
\multicolumn{2}{c}{2$^3$P$_2$: $\Gamma = 113\text{ keV}$} \\
1$^3$S$_1+\gamma$ & 100. \\
\tableline
\multicolumn{2}{c}{3$^3$D$_1$: $\Gamma = 173\text{ keV}$} \\
1$^3$S$_1+\pi\pi$ & 18. \\
2$^3$P$_2+\gamma$ & 2. \\
2$(1^+)+\gamma$ & 29. \\
2$^3$P$_0+\gamma$ & 51. \\
\tableline
\multicolumn{2}{c}{3$^3$D$_2$: $\Gamma = 146\text{ keV}$} \\
1$^3$S$_1+\pi\pi$ & 22. \\
2$^3$P$_2+\gamma$ & 17. \\
2$(1^+)+\gamma$ & 61. \\
\tableline
\multicolumn{2}{c}{3$^3$D$_3$: $\Gamma = 131\text{ keV}$} \\
1$^3$S$_1+\pi\pi$ & 24. \\
2$^3$P$_2+\gamma$ & 76. \\
\tableline
\multicolumn{2}{c}{3$^1$D$_2$: $\Gamma = 124\text{ keV}$} \\
1$^1$S$_0+\pi\pi$ & 26. \\
2$(1^{+\prime})+\gamma$ & 74. \\
\tableline
\multicolumn{2}{c}{3$^3$P$_0$: $\Gamma = 71\text{ keV}$\tablenotemark[1]} \\
2$^3$P$_0+\pi\pi$ & 2. \\
1$^3$S$_1+\gamma$ & 31. \\
2$^3$S$_1+\gamma$ & 57. \\
3$^3$D$_1+\gamma$ & 10. \\
\tableline
\multicolumn{2}{c}{3$(1^+)$: $\Gamma = 86\text{ keV}$\tablenotemark[1]} \\
1$^3$S$_1+\gamma$ & 26. \\
2$^3$S$_1+\gamma$ & 63. \\
3$^3$D$_2+\gamma$ & 11. \\
\tableline
\multicolumn{2}{c}{3$(1^{+\prime})$: $\Gamma = 21\text{ keV}$\tablenotemark[1]}
\\
2$(1^{+\prime})+\pi\pi$ & 7. \\
1$^3$S$_1+\gamma$ & 10. \\
2$^3$S$_1+\gamma$ & 26. \\
3$^3$D$_2+\gamma$ & 55. \\
\tableline
\multicolumn{2}{c}{3$^3$P$_2$: $\Gamma = 122\text{ keV}$\tablenotemark[1]} \\
2$^3$P$_2+\pi\pi$ & 1. \\
1$^3$S$_1+\gamma$ & 21. \\
2$^3$S$_1+\gamma$ & 60. \\
3$^3$D$_3+\gamma$ & 15. \\
3$^3$D$_2+\gamma$ & 3. \\
\end{tabular}
\tablenotemark[1]{Should this state lie above flavor threshold,
dissociation into $BD$ will dominate over the tabulated decay modes.}
\label{totalwidth}
\end{table}

\begin{figure}
\caption{The spectrum of $c\bar{b}$ states.}\label{spectre}
\end{figure}

\begin{figure}
\caption{Normalized dipion mass spectrum for the transition
$2^3\text{S}_1\rightarrow 1^3\text{S}_1 + \pi\pi$ in the $\psi$ (dashed
curve), $B_c$ (solid curve), and $\Upsilon$ (dotted curve) families.}
\label{hadfig}
\end{figure}


\begin{references}
\bibitem{correct} If, for example, we interpret the factor
$(1-16\alpha_s/3\pi)$ as the beginning of an expansion for
$(1+16\alpha_s/3\pi)^{-1}$ with $\alpha_s=0.36$, then the predictions for
the $\psi$ family agree with experiment, within errors, while those for the
$\Upsilon$ family are about $20\%$ low.
\bibitem{detail} It must be remembered in applying these relations to the
$c\bar{b}$ system that the physical eigenstates with $J=\ell\neq 0$
are linear combinations of the equal-mass spin-singlet and spin-triplet
states.
\end{references}
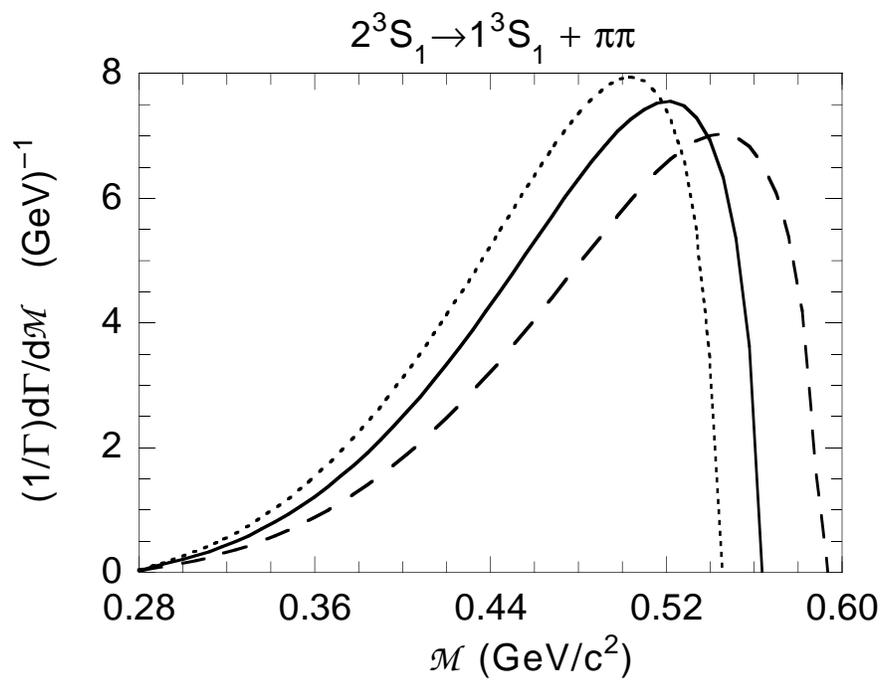
\end{document}